\documentclass[journal ]{new-aiaa}
\usepackage[utf8]{inputenc}
\usepackage{textcomp}

\usepackage{graphicx}
\usepackage{amsmath}
\usepackage[version=4]{mhchem}
\usepackage{siunitx}
\usepackage{longtable,tabularx}
\usepackage{algorithm,algorithmic}
\title{Safety Guaranteed Control for Spacecraft Inspection Mission}

\author{Kun Wang\footnote{Ph.D.Student, School of Aeronautics and Astronautics;wang\_kun@zju.edu.cn}}
\author{Tao Meng\footnote{Professor, School of Aeronautics and Astronautics;mengtao@zju.edu.cn}}
\author{Jiakun Lei\footnote{Ph.D.Student, School of Aeronautics and Astronautics;leijiakun@zju.edu.cn}}
\author{Weijia Wang\footnote{Ph.D.Student, School of Aeronautics and Astronautics;weijiawang@zju.edu.cn}}
\affil{Zhejiang University, Hangzhou, 310027 Zhejiang, People's Republic of China}

\begin{document}
\maketitle

\section{Introduction}
Recently, spacecraft inspection mission has raised significant interest due to its crucial role played in contemporary On-Orbit Service scenario. The Draper Laboratory at the University of California \citep{Day_2020} has considered the potential risks of large-scale spacecraft on-orbit inspection missions and designed a low-risk, low-kinetic, semi-autonomous 3U-sized CubeSat to inspect the outer area of the International Space Station (ISS). Nakka et al. \citep{nakka2022information} proposed a mission scenario in which a large spacecraft releases multiple CubeSats to complete inspections on its surface. Meanwhile, Sepideh et al. \citep{faghihi2023multiple} investigated multiple spacecraft full-coverage inspection missions for the Lunar Gateway.

Notably, in the spacecraft inspection mission described above, the service spacecraft must contend with several safety constraints. Unlike scenarios described in other literature \citep{capolupo2019receding}\citep{fu2022disturbance}, where the target spacecraft is unreasonably treated as a point mass, in this case, the service spacecraft needs to inspect the surface of the larger target spacecraft considering geometric constraints. As such, the obstacle avoidance problem becomes more complex, as multiple components of the target spacecraft must be taken into consideration. Additionally, the optical devices installed on the service spacecraft must be designed to mitigate interference from environmental factors, such as sunlight, to ensure accurate inspections.

Numerous research efforts have been dedicated to addressing spacecraft control under multiple safety constraints. The methods to deal with safety constraints include optimization algorithm based methods, such as model predictive control (MPC)\citep{weiss2015model}\citep{fear2022implementation}\citep{specht2023autonomous} and quadratically constrained quadratic
programming(QCQP) problem\citep{dai2015path}, path planning methods\citep{kjellberg2013discretized}\citep{biggs2016geometric}, and artificial potential function(APF) based methods \citep{wang2022artificial}\citep{hwang2022collision}\citep{lin2022specific}\citep{shao2022fault}. 
Recent advancements in this field include the combination of APF and MPC by Menegatti et al. \citep{menegatti2022model}, who proposed a collision-free formation control strategy for spacecraft formation flying.
Fabio et al.\citep{celani2020spacecraft} proposed a gradient-based attitude motion planning method, which can deal with multiple pointing constraints for attitude rest-to-rest maneuvers. Despite the progress made by these methods, they each have their limitations. MPC methods may suffer from a large computational burden, while APF methods can encounter issues such as getting stuck in local minimum points and difficulties in constructing potential functions for multiple obstacles. Although there have been efforts to address these limitations, existing methods still struggle with safety constraints that involve logical relationships.

The control barrier function (CBF) has emerged as a novel method for safety-critical control. Breeden et al.\citep{breeden2021guaranteed} applied robust control barrier function in spacecraft docking missions in the presence of disturbance. Molnar et al.\citep{molnar2021model} proposed a control strategy that synthesizes the safe velocity based on control barrier function and tracks the safe velocity for safety control, however, the control strategy requires exponential convergence velocity tracking control. A crucial problem in this control strategy is that safety can not be guaranteed because of the velocity tracking error. Singletary et al.\citep{singletary2021safety} proposed a safety-critical paradigm by using energy-based CBF to guarantee safety at the level of dynamics, but system uncertainties are not taken into consideration. 

In this paper, we propose a cascaded safety control strategy for spacecraft inspection missions based on CBF, which is capable of guaranteeing safety in the form of multiple constraints with complex logical relationships. We divide the safety-critical control problem into two parts: safe velocity generation part and velocity tracking part. In the safe velocity generation part, a nominal virtual control law is first designed. The CBF technique is then introduced to be a safety-filter, generating a modification to the nominal virtual control, by solving a quadratic programming (QP) problem. The QP problem can be simple and easy to solve because of the uncertainty-free nature and simple form of the kinematics. A key problem in this strategy is that the actual velocity cannot tracking the safe velocity precisely because of limited control ability of the low-level controller. Thus, in velocity tracking part, proportional-like controllers are designed for position and attitude control, which can guarantee system safety and stability despite the velocity tracking error and relax the performance requirement for necessary fast tracking of the safe velocity. The stability and safety of the controllers are rigorously analyzed in this paper.


This note is organized as follows: Sec.\ref{sec2} introduces spacecraft dynamics and problem formulation. Sec.\ref{sec3} describes attitude and position constraints. Sec.\ref{sec4} presents control strategy design and conducts safety, stability analysis. In Sec.\ref{sec5}, numerical simulation results are shown to illustrate the effectiveness of control strategy. Conclusions are given in Sec.\ref{sec6}.

\section{Spacecraft System Model}
\label{sec2}
\subsection{Necessary Definitions}
In this paper, $\boldsymbol{O}_{n\times n}$, $\boldsymbol{I}_{n\times n}$ denote $n$ order zero matrix and identity matrix respectively, $(\cdot)^{\times}$ denotes the cross-product operator, $\left\| \cdot\right\|$ represents 2-norm of a vector or the corresponding induced matrix norms, $\lambda_{\min}(\cdot)$ means the minimum eigenvalues of a matrix. We define the Earth inertial frame $\mathcal{I}$, the body-fixed frame of the service spacecraft (denoted as the service for brevity) $\mathcal{B}$ and vehicle-velocity-local-horizontal (VVLH) frame of the target (denoted as the target for brevity) $\mathcal{O}$. 

\emph{Definition} 1\citep{ames2019control}:
For the nonlinear affine control system: $\dot{\boldsymbol{x}}=\boldsymbol{f}(\boldsymbol{x})+\boldsymbol{g}(\boldsymbol{x})\boldsymbol{u}$, with $f$ and $g$ locally Lipschitz, $\boldsymbol{x}\in D\subset\mathbb{R}^n$ and $\boldsymbol{u}\in U\subset\mathbb{R}^m$. Let $\mathcal{C}\in D \in \mathbb{R}^n$  be the superlevel set of a continuously differentiable function $h: D\rightarrow \mathbb{R}$, then $h$ is a control barrier function if there exists an extended class $\mathcal{K}_{\infty}$ function $\alpha$ such that for the control system:
\begin{equation*}
	\sup_{\boldsymbol{u}\in U}\left[L_fh(\boldsymbol{x})+ L_gh(\boldsymbol{x})\boldsymbol{u}\right] \geq-\alpha(h(\boldsymbol{x}))
\end{equation*}
for all $\boldsymbol{x}\in D$.
\subsection{Relative Orbit Dynamics}
Let $\boldsymbol{r}$, $\boldsymbol{v}$ be the relative position and velocity vector from the target to the service expressed in frame $\mathcal{O}$ respectively, $m$ is the mass of the service, $\boldsymbol{F}$ denotes the position control force in frame $\mathcal{O}$, $g=\mu(\boldsymbol{r}_t/\left\| \boldsymbol{r}_s\right\|^3-\boldsymbol{r}_t/\left\| \boldsymbol{r}_t\right\|^3 )$, $\boldsymbol{r}_t$ and $\boldsymbol{r}_s$ denote the position vectors of the serve and the target in frame $\mathcal{I}$ respectively. Then the relative orbit dynamics between the service and the target can be written as \citep{dong2017safety}
\begin{subequations}
	\label{eq1}
	\begin{align}
		\dot{\boldsymbol{r}}&=\boldsymbol{v} \label{eq1a} \\
		\dot{\boldsymbol{v}}&=-\boldsymbol{C}_o\boldsymbol{v}-\boldsymbol{D}_o\boldsymbol{r}+\boldsymbol{g}+\frac{\boldsymbol{F}}{m} + \boldsymbol{d}_f\label{eq1b}
	\end{align}
\end{subequations}
where $\boldsymbol{C}_o=\begin{bmatrix}0& 2\dot{f}_{\theta} &0 \\-2\dot{f}_{\theta}& 0 &0 \\0& 0 &0\end{bmatrix}$, $\boldsymbol{D}_o=\begin{bmatrix}\dot{f}_{\theta}^2& \ddot{f}_{\theta} &0 \\-\ddot{f}_{\theta}& \dot{f}_{\theta}^2 & 0\\0& 0 &0\end{bmatrix}$, $\boldsymbol{d}_f$ denotes external disturbance and unmodeled dynamics. $f_{\theta}$ is the true anomaly of the target, and for arbitrary orbit, $\dot f_{\theta}  = \sqrt {\frac{\mu }{{{a^3}{{(1 - {e^2})}^3}}}} {\left(1 + e\cos (f_{\theta} )\right)^2}$, $\ddot f_{\theta}  =  - \frac{{2\mu e\sin f_{\theta} {{(1 + e\cos (f_{\theta} ))}^3}}}{{{a^3}{{(1 - {e^2})}^3}}}$. $a$ denotes the semimajor axis of the target, $e$ denotes the eccentricity of the target, $\mu$ is the gravitational constant.
\subsection{Attitude Dynamics}
In this paper, we use the reduced attitude control model in the following analysis. Let $\boldsymbol{\Gamma}_{\mathcal{I}}\in\mathbb{S}^2=\left\lbrace\boldsymbol{b}\in \mathbb{R}^3|\boldsymbol{b}^T\boldsymbol{b}=1\right\rbrace$ be the desired direction in frame $\mathcal{I}$, and vector $\boldsymbol{\Gamma}$ is the expression of $\boldsymbol{\Gamma}_{\mathcal{I}}$ in frame $\mathcal{B}$. Then we have $\boldsymbol{\Gamma}=\boldsymbol{R}^T\boldsymbol{\Gamma}_{\mathcal{I}}$, where $\boldsymbol{R}$ is the rotation matrix from frame $\mathcal{B}$ to frame $\mathcal{I}$. Besides, ${\boldsymbol{\omega}}$ denotes the angular velocity vector expressed in frame $\mathcal{B}$, $\boldsymbol{J}$ denotes the inertia of the service and $\boldsymbol{T}$ denotes the attitude control torque of the service. 
 Then the attitude dynamics can be expressed as \citep{chaturvedi2011rigid}
\begin{subequations}
	\label{eq2}
	\begin{align}
		\dot{\boldsymbol{\Gamma}}&=\boldsymbol{\Gamma}\times\boldsymbol{\omega} \label{eq2a}\\
		{\dot{\boldsymbol{\omega}}}&=\boldsymbol{J}^{-1}\left[ -{\boldsymbol{\omega}}\times(\boldsymbol{J}{\boldsymbol{\omega}})+\boldsymbol{T}\right] +\boldsymbol{d}_t \label{eq2b}
	\end{align}
\end{subequations}
in which $\boldsymbol{d}_t$ denotes external disturbance and unmodeled dynamics.


\emph{Assumption} 1: The service can be regarded as a mass point in the safety control, which means that the envelop of the service is small enough to be ignored compared with the target.

\emph{Assumption} 2: The initial states of the service satisfy the safe constraints and the desired states are also in the safe region.

\emph{Assumption} 3: The derivatives of the terms $\boldsymbol{d}_f$ and $\boldsymbol{d}_t$ are bounded, i.e.$\left\|\dot{\boldsymbol{d}} \right\| \leq \delta$, where $\boldsymbol{d}=\left[ \boldsymbol{d}_f^T,\boldsymbol{d}_t^T\right]^T$.

\section{Constraints Description and Problem Formulation}
\label{sec3}
\subsection{Position Constraints}
\begin{figure}[htb]
	\centering
	\includegraphics[scale=0.2]{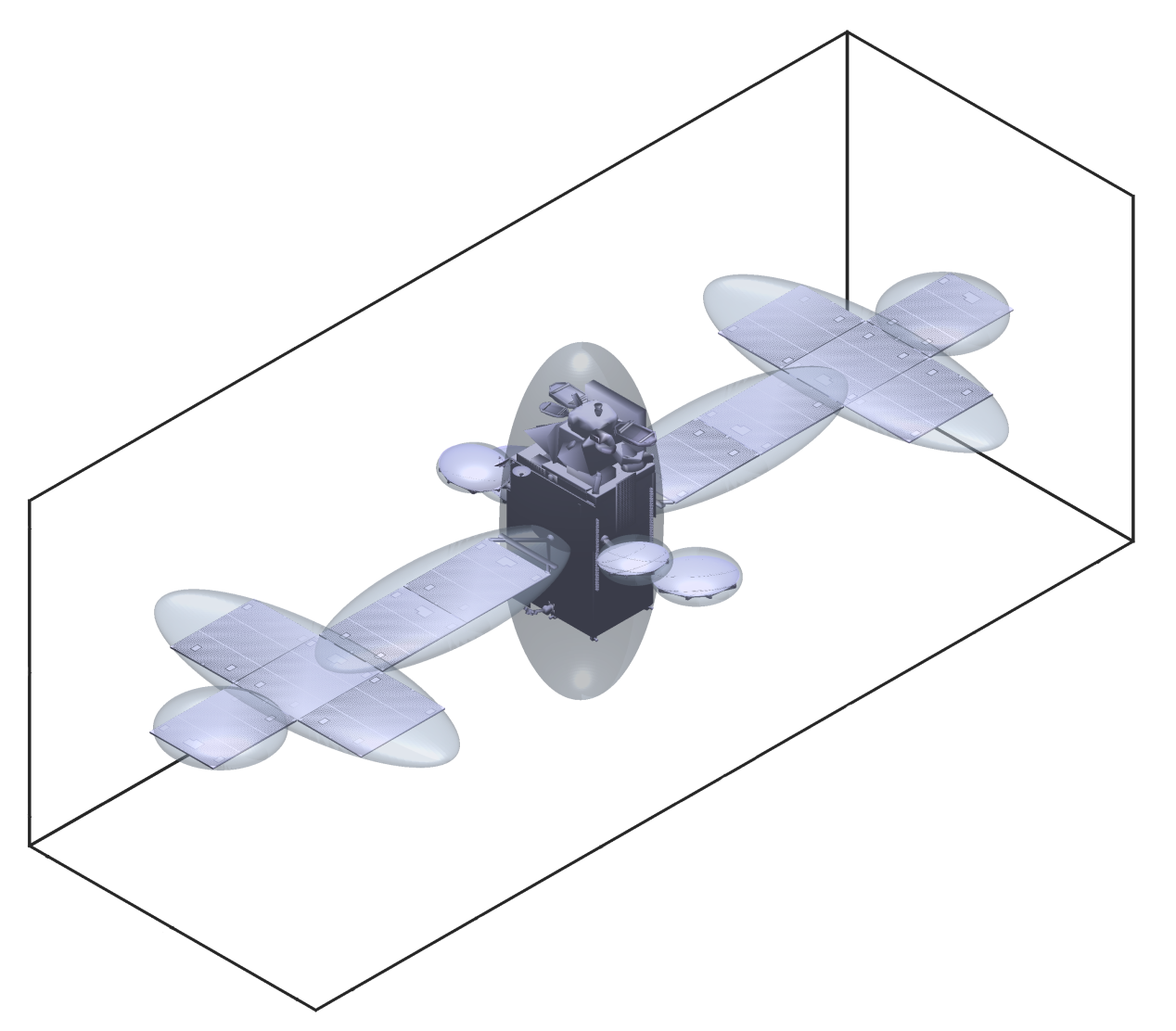}
	\caption{Model of the target spacecraft}
	\label{fig1}
\end{figure}

The position constraints of the service mainly reflected in avoiding collisions with the target. The target's body and its attached solar panels, antennas and other components are modeled as ellipsoids of different size, that is, obstacles are wrapped with ellipsoids and the inside of ellipsoids is recorded as unsafe areas, as Fig\ref{fig1} shows.

Let $\boldsymbol{r}=\left[ r_x,r_y,r_z\right] ^T$ be the current position coordinates of the service. $(a_i,b_i,c_i)$ denotes the center point of $i$-th obstacle ($i=1,2,\cdots,m$), $(l_{xi},l_{yi},l_{zi})$ denotes the length, width and height of the $i$-th ellipsoid envelope. The control barrier function of the $i$-th obstacle can be written as
\begin{equation}
	\label{eq3a}
		h_{pi}(\boldsymbol{r})=\left(\frac{r_x-a_i}{l_{xi}}\right)^2+\left(\frac{r_y-b_i}{l_{yi}}\right)^2+\left(\frac{r_z-c_i}{l_{zi}}\right)^2-1 
\end{equation}
and the partial derivative of $h_{pi}$ to $\boldsymbol{r}$ is
\begin{equation}
	\label{eq3b}
	\boldsymbol{Y}_{pi}=\frac{\partial h_{pi}}{\partial \boldsymbol{r}}=\begin{bmatrix}\frac{2}{l_{xi}^2}(r_x-a_i)&\frac{2}{l_{yi}^2}(r_y-b_i)&\frac{2}{l_{zi}^2}(r_z-c_i)\end{bmatrix}
\end{equation}

When $h_{pi}\geq0$, the service is outside of the obstacle, which means that the position constraint is satisfied.
\subsection{Attitude Constraints}

The attitude constraints of the service mainly come from the interference of the sunlight in space and the reflected light from the Earth on the optical devices. For example, the star tracker will fail to provide useful attitude information when it is interfered by the light from the Sun or the Earth, and some cameras will not allow sunlight to enter a certain field of view. These constraints can be summarized as: the angle between the optical axis of the optical device and the space environment vector (such as sun vector, earth vector etc.) should be smaller than the safety threshold.
\begin{figure}[hbt!]
	\centering
	\includegraphics[scale=0.69]{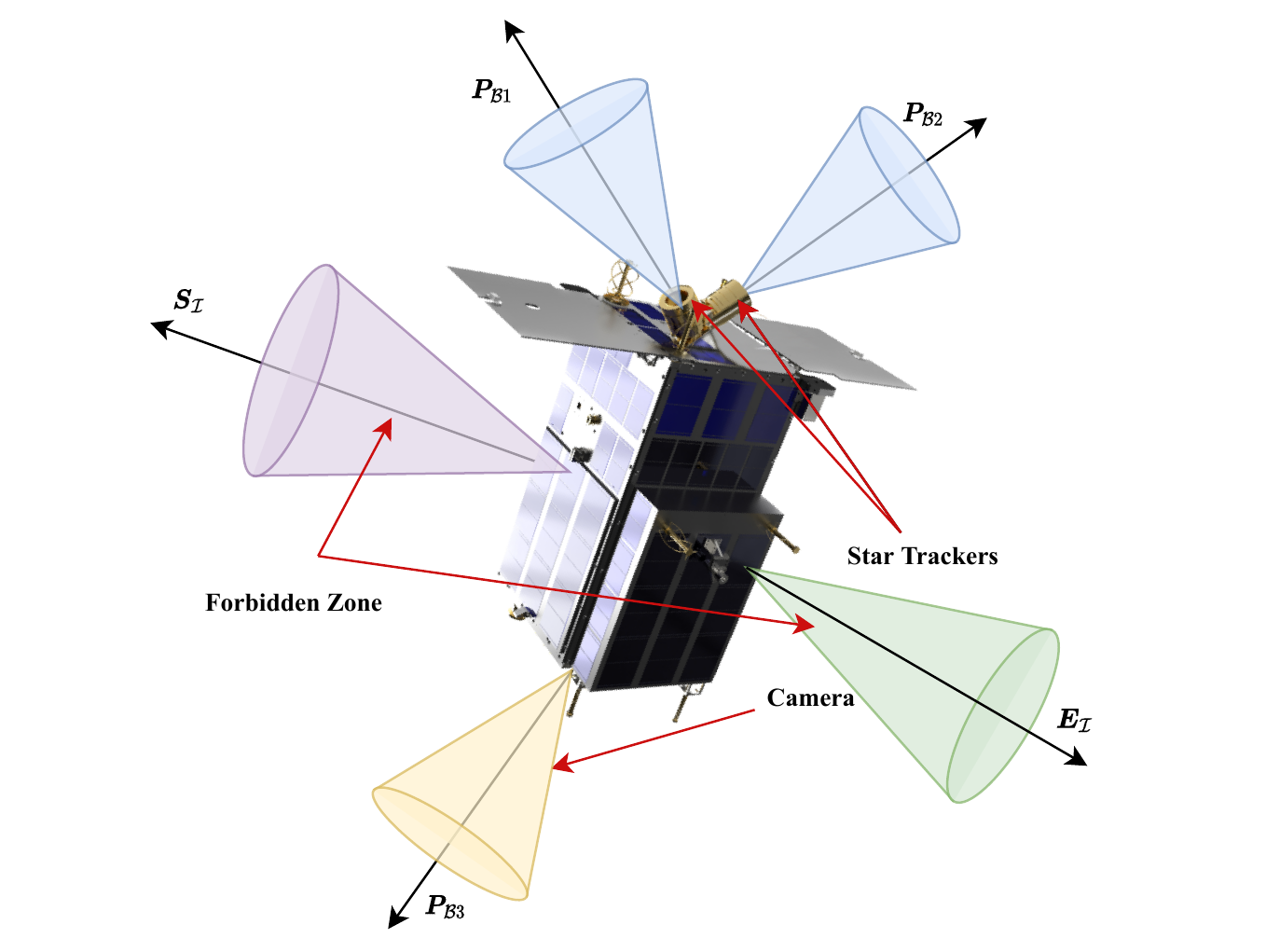}
	\caption{Illustration of the service's attitude constraints}
	\label{fig2}
\end{figure}

In this paper, as Fig.(\ref{fig2})shows, we consider a small satellite equipped with a camera and two star trackers. $\boldsymbol{S}_{\mathcal{I}}$ and $\boldsymbol{E}_{\mathcal{I}}$ denote the forbidden zone respectively. The sun exclusive angles of the star tracker and the camera are denoted as $\Psi_{s}$ and $\Psi_{c}$ respectively, and the Earth exclusive angle for star tracker is denoted as $\Psi_{e}$. Let $\boldsymbol{P}_{\mathcal{B}i}$ and $\boldsymbol{P}_{\mathcal{I}i}$ denote the unit direction vector of the $i$-th optical axis ($i=1,2,3$, corresponding to star tracker 1, star tracker 2 and the camera) expressed in frame $\mathcal{B}$ and frame $\mathcal{I}$ respectively, then we have $\boldsymbol{P}_{\mathcal{I}i}=\boldsymbol{R}\boldsymbol{P}_{\mathcal{B}i}$. Let $\boldsymbol{V}_{\mathcal{I}s}$ and $\boldsymbol{V}_{\mathcal{I}e}$ denote the unit direction of the sun vector and the unit direction of the Earth vector in frame $\mathcal{I}$ respectively. Then the attitude constraints of the service can be written as follows:
\begin{subequations}
	\label{eq8}
	\begin{align}
		h_{a1}=\cos \Psi_{s} - \boldsymbol{V}_{\mathcal{I}s}^T\boldsymbol{R}\boldsymbol{P}_{\mathcal{B}1}  \\
		h_{a2}=\cos \Psi_{e} - \boldsymbol{V}_{\mathcal{I}e}^T\boldsymbol{R}\boldsymbol{P}_{\mathcal{B}1} \\
		h_{a3}=\cos \Psi_{s} - \boldsymbol{V}_{\mathcal{I}s}^T\boldsymbol{R}\boldsymbol{P}_{\mathcal{B}2} \\
		h_{a4}=\cos \Psi_{e} - \boldsymbol{V}_{\mathcal{I}e}^T\boldsymbol{R}\boldsymbol{P}_{\mathcal{B}2} \\
		h_{a5} =\cos \Psi_{c} - \boldsymbol{V}_{\mathcal{I}s}^T\boldsymbol{R}\boldsymbol{P}_{\mathcal{B}3} \label{eq8e}
	\end{align}
\end{subequations} 
where $h_{a1}$, $h_{a2}$ and $h_{a3}$, $h_{a4}$ denote the CBFs of the two star trackers constrained by sun vector and the Earth vector respectively, $h_{a5}$ denotes the camera constraint. Taking the time derivative of $h_{a5}$ yields
\begin{equation}
	\label{eq9}
	\dot{h}_{a5}=-\boldsymbol{V}_{\mathcal{I}s}^T\boldsymbol{R}\boldsymbol{\omega}_s^{\times}\boldsymbol{P}_{\mathcal{B}3}=\boldsymbol{V}_{\mathcal{I}s}^T\boldsymbol{R}\boldsymbol{P}_{\mathcal{B}3}^{\times}\boldsymbol{\omega}_s
\end{equation}

We denote $\boldsymbol{Y}_{a5}=\boldsymbol{V}_{\mathcal{I}s}^T\boldsymbol{R}\boldsymbol{P}_{\mathcal{B}3}^{\times}$ as the Lie-like derivative definition for Eq.(\ref{eq8e}) and other equations in Eq.(\ref{eq8}) have the same definition.

Different from position constraints, there will be some logical relationships between attitude constraints. We use $\vee$ and $\wedge$ to describe the "or" and "and" logic. The control strategy needs to ensure that at least one star tracker is available for attitude determination and the constraint of the camera also needs to be satisfied at the same time. Thus the logical relationships between these CBFs can be written as

\begin{equation}
	\label{eq10}
	h_{\min}\triangleq((h_{a1}\wedge h_{a2})\vee(h_{a3}\wedge h_{a4}))\wedge h_{a5}
\end{equation}

From Eq.(\ref{eq10}), we can know that the attitude constraints are satisfied when $h_{\min}\geq0$.

\subsection{Control Objective}
The objective of this paper is to design a control strategy for the service to accomplish inspection missions safely. We consider the commercial communications satellite Intelsat-30 as the target and a small satellite with optical equipment as the service. The service needs to navigate to designated checkpoints sequentially and observe the target from a specific pose. Moreover, strict safety constraints on attitude and position must be adhered to throughout the mission.
\section{Control Design and Safety, Stability Analysis}
\label{sec4}
In this section, a disturbance observer is designed firstly to deal with the model uncertainty and external disturbance. After that, safe velocity and safe angular velocity are generated. Then proportional-like attitude and position controllers are designed. The safety and stability analysis are conducted after controller design.
\subsection{Disturbance Observer Design}
Following the results of \citep{li2014disturbance}, we design a disturbance observer. Define the velocity state $\boldsymbol{x}=\left[\boldsymbol{v}^T,\boldsymbol{\omega}^T \right] ^T$, then the time derivative of $\boldsymbol{x}$ can written as follows according to Eq.(\ref{eq2b}) and Eq.(\ref{eq3b})
\begin{equation}
	\dot{\boldsymbol{x}}=\boldsymbol{M}(\boldsymbol{x})+\boldsymbol{N}(\boldsymbol{x})\boldsymbol{u}+\boldsymbol{d}
	\label{eq15}
\end{equation}
where $\boldsymbol{u}=\left[ \boldsymbol{F}^T,\boldsymbol{T}^T\right]^T $, $\boldsymbol{d}=\left[ \boldsymbol{d}_f^T,\boldsymbol{d}_t^T\right]^T $, $\boldsymbol{M}(\boldsymbol{x})=\begin{bmatrix}-\boldsymbol{C}_o\boldsymbol{v}-\boldsymbol{D}_o\boldsymbol{r}+\boldsymbol{g}\\ -\boldsymbol{J}^{-1}\left( \boldsymbol{\omega}\times(\boldsymbol{J\omega})\right) \end{bmatrix}$, $\boldsymbol{N}(\boldsymbol{x})=\begin{bmatrix}\frac{1}{m}\boldsymbol{I}_{3\times3}& \\& \boldsymbol{J}^{-1}\end{bmatrix}$.

Then the lumped disturbance $\boldsymbol{d}$ can be estimated by the following observer:
\begin{equation}
	\label{eq16}
	\begin{aligned}
		\hat{\boldsymbol{d}}&=\boldsymbol{z}+\boldsymbol{p}(\boldsymbol{x})\\
		\dot{\boldsymbol{z}}&=-\boldsymbol{L}\boldsymbol{z}-\boldsymbol{L}\left[ \boldsymbol{p}(\boldsymbol{x})+\boldsymbol{M}(\boldsymbol{x})+\boldsymbol{N}(\boldsymbol{x})\boldsymbol{u}\right] 
	\end{aligned}
\end{equation}
where $\hat{\boldsymbol{d}}=\left[\boldsymbol{d}_f^T,\boldsymbol{d}_t^T \right]^T$, $\boldsymbol{L}=\frac{\partial \boldsymbol{p}(\boldsymbol{x})}{\partial \boldsymbol{x}}$ is a positive definite matrix and satisfies $\lambda_{\min}(\boldsymbol{L})>\frac{\mu}{2}$, $\mu$ is a constant to be selected. Thus, $\boldsymbol{p}(\boldsymbol{x})$ can be designed as $\boldsymbol{p}(\boldsymbol{x})=\boldsymbol{Lx}$.
 
\emph{Theorem} 1: Under the assumption above, the disturbance estimation error for the system dynamics Eq.(\ref{eq15}) will be uniformly ultimately bounded with the disturbance observer Eq.(\ref{eq16}).

\emph{Proof}: The proof is relegated to the Appendix A.
\subsection{Position Control}
\subsubsection{Safe Velocity Generation}
Denote the desired inspection position as $\boldsymbol{r}_d$ and let $\boldsymbol{\dot{r}}_d=0$. $\boldsymbol{r}_e=\boldsymbol{r}-\boldsymbol{r}_d$ is the position control error. 
Let the virtual position control variable be
\begin{equation}
	\label{eq6}
	\boldsymbol{v}_c=-k_{p1}\boldsymbol{r}_e
\end{equation}
where $k_{p1}>0$. Let $\boldsymbol{v}_s$ be the safe velocity that satisfies all of the position constraints, then $\boldsymbol{v}_s$ can be obtained by solving the following QP problem (QP1):
\begin{subequations}
	\label{eq7}
	\begin{align}
		\arg &\min_{\boldsymbol{v}_s}\left\|\boldsymbol{v}_s-\boldsymbol{v}_c \right\| \\
		&s.t. \quad \boldsymbol{v}_{\min}\leq\boldsymbol{v}_s\leq\boldsymbol{v}_{\max}\label{eq7b}\\
		&\quad \boldsymbol{Y}_{pi}\boldsymbol{v}_s\geq-\alpha_{pi}(h_{pi})+\gamma_p\left\|\boldsymbol{Y}_{pi} \right\| \label{eq7c}
	\end{align}
\end{subequations}
where $\alpha_{pi}$ is an extended class $\mathcal{K}_{\infty}$ function, $\gamma_p$ is a constant related to the disturbance estimation error, which will be explained in Sec.\ref{sec4}.

\emph{Remark} 1: The term $\gamma_p\left\|\boldsymbol{Y}_{pi} \right\|$ adds the conservatism of CBF, introduced by the disturbance observer. It is necessary to guarantee strict safety in the presence of disturbance, although the addition of this residual term increases the conservatism of the CBF.
\subsubsection{Position Controller Design}
Let the extended class  $\mathcal{K}_{\infty}$ function for the position CBFs be selected as $\alpha_{p1}(\cdot)=\alpha_{p2}(\cdot)=\cdots=\alpha_{pm}(\cdot)=\alpha_p$ and the position controller can be designed as
\begin{equation}
	\label{eq17}
	\boldsymbol{F}=m\left[\underbrace{\boldsymbol{C}_o\boldsymbol{v}+\boldsymbol{D}_o\boldsymbol{r}-\boldsymbol{g}}_{(1)}\underbrace{-\left(\alpha_p+k_{p2}\right)\boldsymbol{v}+k_{p2}\boldsymbol{v}_s}_{(2)}\underbrace{-\hat{\boldsymbol{d}}_f}_{(3)}\right]
\end{equation}
where $\alpha_p>0$ and $k_{p2}>0$ are the control parameters to be designed. As Eq.(\ref{eq17}) shows, the position controller can be divided into three terms: (1) the feedforward term to compensate the system dynamics; (2) the safety correction term; (3) the rejection term to external disturbance and unmodeled dynamics.
\subsubsection{Safety Analysis for Position Controller}
Define the following position safety function
\begin{equation}
	\label{eq18}
	B_{pi}(\boldsymbol{r},\boldsymbol{v})=\boldsymbol{Y}_{pi}\boldsymbol{v}+\alpha_ph_{pi}(\boldsymbol{r}), \quad i=1,2,\cdots,m
\end{equation}

For the relative orbit dynamics Eq.(\ref{eq1b}), if the initial value of $B_{pi}(\boldsymbol{r}(0),\boldsymbol{v}(0))$ and $h_{pi}(0)$ are non-negative, then all the position constraints will be always satisfied under the position controller Eq.(\ref{eq17}), i.e. $B_{pi}$ is a CBF. Here detailed description of the safety is presented.

Taking the time derivative of Eq.(\ref{eq18}) yields
\begin{equation}
	\label{eq19}
	\dot{B}_{pi}(\boldsymbol{r},\boldsymbol{v})=\dot{\boldsymbol{Y}}_{pi}\boldsymbol{v}+\boldsymbol{Y}_{pi}{\dot{\boldsymbol{v}}}+\alpha_p\boldsymbol{Y}_{pi}\boldsymbol{v} \\
\end{equation}

Considering Eq.(\ref{eq3b}) and substitute Eq.(\ref{eq2b}) into Eq.(\ref{eq19}) yields

\begin{equation}
	\label{eq20}
	\dot{B}_{pi}(\boldsymbol{r},\boldsymbol{v})=\boldsymbol{v}^T\boldsymbol{P}\boldsymbol{v}+\boldsymbol{Y}_{pi}\left[-\boldsymbol{C}_o \boldsymbol{v}-\boldsymbol{D}_o\boldsymbol{r}+\boldsymbol{g}+\frac{\boldsymbol{F}}{m}+\boldsymbol{d}_f\right] +\alpha_p\boldsymbol{Y}_{pi}\boldsymbol{v}
\end{equation}
where $\boldsymbol{P}=\operatorname{diag}\left(\frac{2}{l_{xi}^2},\frac{2}{l_{yi}^2},\frac{2}{l_{zi}^2} \right)$ is a positive definite matrix, thus $\boldsymbol{v}^T\boldsymbol{P}\boldsymbol{v}\geq0$. 

Then substituting Eq.(\ref{eq17}) into Eq.(\ref{eq20}) yields
\begin{equation}
	\label{eq21}
	\begin{aligned}
		\dot{B}_{pi}(\boldsymbol{r},\boldsymbol{v})&\geq \boldsymbol{Y}_{pi}\left[-\left(\alpha_p+k_{p2}\right)\boldsymbol{v}+k_{p2}\boldsymbol{v}_s+\boldsymbol{d}_f-{\hat{\boldsymbol{d}}}_f \right] +\alpha_p\boldsymbol{Y}_{pi}\boldsymbol{v} \\
		&= -k_{p2}\boldsymbol{Y}_{pi}\boldsymbol{v}+k_{p2}\boldsymbol{Y}_{pi}\boldsymbol{v}_s+\boldsymbol{Y}_{pi}\boldsymbol{e}_f
	\end{aligned}
\end{equation}

Substituting Eq.(\ref{eq7c}) into Eq.(\ref{eq21}) yields
\begin{equation}
	\label{eq22}
	\begin{aligned}
		\dot{B}_{pi}(\boldsymbol{r},\boldsymbol{v})&\geq-k_{p2}\boldsymbol{Y}_{pi}\boldsymbol{v}-k_{p2}\alpha_ph_{pi}+\gamma_p\left\|\boldsymbol{Y}_{pi}\right\|-\left\|\boldsymbol{Y}_{pi}\right\|\left\|\boldsymbol{e}_f\right\|  \\
		&=-k_{p2}B_i(\boldsymbol{r},\boldsymbol{v})+\gamma_p\left\|\boldsymbol{Y}_{pi}\right\|-\left\|\boldsymbol{Y}_{pi}\right\|\left\|\boldsymbol{e}_f\right\|
	\end{aligned}
\end{equation}

The disturbance estimation error is bounded according to Theorem 1 and we can select a reasonable parameter $\gamma_p$ to make sure that $\gamma_p\geq\left\|\boldsymbol{e}_f\right\|_{\max}$. Then it can be induced that $\dot{B}_{pi}(\boldsymbol{r},\boldsymbol{v})\geq-k_{p2}B_{pi}(\boldsymbol{r},\boldsymbol{v})$, so $B_{pi}(\boldsymbol{r},\boldsymbol{v})\geq0$. Thus, $\boldsymbol{Y}_{pi}\boldsymbol{v}\geq-\alpha_ph_{pi}(\boldsymbol{r})$ holds for all the position constraints, i.e.$\dot{h}_{pi}\geq-\alpha_ph_{pi}$, which means $h_{pi}\geq0$ always holds. As a result, the safety can be guaranteed with the proposed position control strategy. 

\emph{Remark} 2: From above discussion, we can conclude that the proposed controller can achieve safety control without exponential convergence tracking performance requirement. Additionally, the proposed control strategy can be generalized to a class of Euler-Lagrange system collision-free control, in which obstacles can be shaped as spheres or ellipsoids.

\subsubsection{Stability Analysis for Position Controller}
\label{secpsa}
Stability is another important factor for control strategy design. Let $\boldsymbol{v}_e=\boldsymbol{v}-\boldsymbol{v}_c$ be the velocity tracking error in the sense of stability. It should be noted that the velocity tracking error refers to the error between the safe and the actual velocity in other sections, here, we use $\boldsymbol{v}_e$ for stability analysis. In this control strategy, the safe velocity $\boldsymbol{v}_s$ can be regarded as a modification of the virtual velocity $\boldsymbol{v}_c$, thus $\boldsymbol{v}_s$ can be expressed as $\boldsymbol{v}_s=\boldsymbol{v}_c+\Delta \boldsymbol{v}$. Considering the velocity contraints Eq.(\ref{eq7b}), $\Delta \boldsymbol{v}$ is a bounded term.

\emph{Theorem} 2: For the relative orbit dynamics Eq.(\ref{eq1b}), the control error will converge into a compact set, which is related to the modification effect of the safe velocity to the virtual control variable, under the position controller Eq.(\ref{eq17}).

\emph{Proof}: The proof is relegated to the Appendix B.

\emph{Remark} 3: The size of the domain of the convergence is related to the modification effect from QP1, which means that the position control error can converge into a fixed and small set, as the service bypasses the obstacle and $\Delta\boldsymbol{v}$ tends to zero.
\subsection{Attitude Control}
\subsubsection{Safe Angular Velocity Generation}
\label{sec3d}
Let the extended class $\mathcal{K}_{\infty}$ function for attitude CBFs be $\alpha(\cdot)=\alpha_{a}$. 
Define $\boldsymbol{\Phi}^T=\boldsymbol{P}_{\mathcal{B}3}^T\boldsymbol{\Gamma}^{\times}$ and let the virtual control variable be
\begin{equation}
	\label{eq14}
	\boldsymbol{\omega}_c=k_{a1}\boldsymbol{\Phi}
\end{equation}
where $k_{a1}>0$. Let $\boldsymbol{\omega}_s$ be the safe angular velocity, then $\boldsymbol{\omega_s}$ can be obtained from the following steps:

1) \textbf{Step}1: calculate the value of CBFs: $h_{a1}$, $h_{a2}$, $h_{a3}$, $h_{a4}$, $h_{a5}$;

2) \textbf{Step}2: let $h_1=\min \left\lbrace h_{a1},h_{a2}\right\rbrace$, $h_2=\min \left\lbrace h_{a3},h_{a4}\right\rbrace$, $h_3=\max \left\lbrace h_1,h_2\right\rbrace$, $h_4=\min \left\lbrace h_3,h_{a5}\right\rbrace$

3) \textbf{Step}3: set $\boldsymbol{H}_a$ as an empty set;

3) \textbf{Step}4: if $\left| h_4-h_3 \right| \leq \epsilon$, then if $h_3$ is equal to $h_1$, push $h_{a1}$ and $h_{a2}$ into $\boldsymbol{H}_a$, otherwise push $h_{a3}$ and $h_{a4}$ into $\boldsymbol{H}_a$;

4) \textbf{Step}5: if $\left| h_4-h_{a5} \right| \leq \epsilon$, then push $h_{a5}$ into $\boldsymbol{H}_a$;

5) \textbf{Step}6: solve the following QP problem (QP2):
\begin{subequations}
	\label{eqqp2}
	\begin{align}
		\arg &\min_{\boldsymbol{\omega}_s}\left\|\boldsymbol{\omega}_s-\boldsymbol{v}_c \right\| \\
		&s.t. \quad \boldsymbol{\omega}_{\min}\leq\boldsymbol{\omega}_s\leq\boldsymbol{\omega}_{\max}\\
		&\quad \boldsymbol{Y}_{ai}\boldsymbol{\omega}_s\geq-\alpha_{a}h_{ai}+\left\|\boldsymbol{\omega}\right\|^2\sin \theta_i+\gamma_a\left\|\boldsymbol{Y}_{ai} \right\|, \forall h_{ai} \in \boldsymbol{H}_a \label{eq16c}
	\end{align}
\end{subequations}
where $\sin \theta_i$ is the angle between $\boldsymbol{\omega}$ and $\boldsymbol{P}_{\mathcal{B}i}$.

\emph{Remark} 4: The attitude constraints are composed by the boolean operator (see Eq.(\ref{eq10})), which is called boolean nonsmooth control barrier function (BNCBF) defined in \citep{glotfelter2018boolean}. According to Theorem 3 and experimental results in \cite{glotfelter2018boolean}, the safe angular velocity can be generated by using the almost-active gradient as the QP2 problem's constraints.
\subsubsection{Attitude Controller Design}
The attitude controller can be designed as
\begin{equation}
	\label{eq28}
	\boldsymbol{T}=\underbrace{\boldsymbol{\omega}\times(\boldsymbol{J}\boldsymbol{\omega})}_{(1)}+\boldsymbol{J}\left[\underbrace{-\left(\alpha_a+k_{a2}\right)\boldsymbol{\omega}+k_{a2}\boldsymbol{\omega}_s}_{(2)}\underbrace{-\hat{\boldsymbol{d}}_t}_{(3)}\right]
\end{equation}
where $k_{a2}>0$ is the control parameter to be designed. As Eq.(\ref{eq28}) shows, the attitude controller can also be divided into three terms as same as the position controller.
\subsubsection{Safety Analysis for Attitude Controller}
Define the following attitude safety function
\begin{equation}
	\label{eq29}
	B_{ai}(\boldsymbol{\omega})=\boldsymbol{Y}_{ai}\boldsymbol{\omega}+\alpha_a h_{ai},\quad \forall h_{ai}\in \boldsymbol{H}_a
\end{equation}

For the attitude dynamics Eq.(\ref{eq2b}), if the initial value of $B_{ai}(\boldsymbol{\omega}(0))$ and $h_{ai}(0)$ are non-negative, then all the attitude constraints in $\boldsymbol{H}_a$ will be always satisfied under the attitude controller Eq.(\ref{eq28}), i.e.$B_{ai}$ is a CBF. Here detailed discussion about the safety is presented.

Taking the time derivative of Eq.(\ref{eq29}) and substituting Eq.(\ref{eq2}) yields
\begin{equation}
	\label{eq30}
	\begin{aligned}
		\dot{B}_{ai}(\boldsymbol{\omega})&=\dot{\boldsymbol{Y}}_{ai}\boldsymbol{\omega}+\boldsymbol{Y}_{ai}\dot{\boldsymbol{\omega}}+\alpha_a \dot{h}_{ai} \\
		&=\dot{\boldsymbol{Y}}_{ai}\boldsymbol{\omega}+\boldsymbol{Y}_{ai}\left\lbrace \boldsymbol{J}^{-1}\left[ -{\boldsymbol{\omega}}\times(\boldsymbol{J}{\boldsymbol{\omega}})+\boldsymbol{T}\right] +\boldsymbol{d}_t \right\rbrace +\alpha_a \dot{h}_{ai}
	\end{aligned}
\end{equation}

The term in Eq.(\ref{eq30}) $\dot{\boldsymbol{Y}}_{ai}\boldsymbol{\omega}$ can be induced as
\begin{equation}
	\label{31}
	\begin{aligned}
		\dot{\boldsymbol{Y}}_{ai}\boldsymbol{\omega}&=\boldsymbol{V}_{\mathcal{I}j}^T\boldsymbol{R}\boldsymbol{\omega}^{\times}\boldsymbol{P}_{\mathcal{B}i}^{\times}\boldsymbol{\omega} \\
		&=\boldsymbol{V}_{\mathcal{I}j}^T\boldsymbol{R}\left[\boldsymbol{\omega}\times\left(\boldsymbol{P}_{\mathcal{B}i}\times\boldsymbol{\omega}\right)\right]
	\end{aligned}
\end{equation}

Then we have
\begin{equation}
	\label{eq32}
	\begin{aligned}
		\left\|\dot{\boldsymbol{Y}}_{ai}\boldsymbol{\omega}\right\|&=\left\|\boldsymbol{V}_{\mathcal{I}j}^T\boldsymbol{R}\left[\boldsymbol{\omega}\times\left(\boldsymbol{P}_{\mathcal{B}i}\times\boldsymbol{\omega}\right)\right]\right\| \\
		&\leq\left\|\boldsymbol{V}_{\mathcal{I}j}^T\right\|\left\|\boldsymbol{R}\right\|\left\|\boldsymbol{\omega}\times\left(\boldsymbol{P}_{\mathcal{B}i}\times\boldsymbol{\omega}\right)\right\|  \\
		&=\left\|\boldsymbol{\omega}\right\| \left\|\boldsymbol{P}_{\mathcal{B}i}\times\boldsymbol{\omega}\right\|
		=\left\|\boldsymbol{\omega}\right\|^2\sin \theta_i
	\end{aligned}
\end{equation}
where $\sin \theta_i$ is the angle between $\boldsymbol{\omega}$ and $\boldsymbol{P}_{\mathcal{B}i}$. Substitute Eq.(\ref{eq28}) into Eq.(\ref{eq30}) and then Eq.(\ref{eq30}) can be induced as
\begin{equation}
	\label{eq33}
	\begin{aligned}
		\dot{B}_{ai}(\boldsymbol{\omega})&\geq -\left\|\boldsymbol{\omega}\right\|^2\sin \theta_i+\boldsymbol{Y}_{ai}\left[-\left(\alpha_a+k_{a2}\right)\boldsymbol{\omega}+\boldsymbol{\omega}_s+\boldsymbol{d}-\hat{\boldsymbol{d}}_t\right] +\alpha_a \dot{h}_{ai}\\
		&=-\left\|\boldsymbol{\omega}\right\|^2\sin \theta_i-(\alpha_a+k_{a2})\boldsymbol{Y}_{ai}\boldsymbol{\omega}+\boldsymbol{Y}_{ai}\boldsymbol{\omega_s}+\boldsymbol{Y}_{ai}\boldsymbol{e}_t+\alpha_a \dot{h}_{ai}\\
		&\geq-\left\|\boldsymbol{\omega}\right\|^2\sin \theta_i-k_{a2}\boldsymbol{Y}_{ai}\boldsymbol{\omega}+\boldsymbol{Y}_{ai}\boldsymbol{\omega_s}-\left\|\boldsymbol{Y}_{ai}\right\|\left\|\boldsymbol{e}_t\right\|
	\end{aligned}
\end{equation}

Substituting Eq.(\ref{eq16c}) into Eq.(\ref{eq33}) yields
\begin{equation}
	\label{eq34}
	\begin{aligned}
		\dot{B}_{ai}(\boldsymbol{\omega})&\geq -k_{a2}\boldsymbol{Y}_{ai}\boldsymbol{\omega}-\alpha_a h_{ai}+\left\|\boldsymbol{Y}_{ai}\right\| \left(\gamma_a-\left\|\boldsymbol{e}_t\right\|\right) 
	\end{aligned}
\end{equation}

As the disturbance estimation error is bounded, we can select a proper $\gamma_a$ to satisfy $\gamma_a>\left\|\boldsymbol{e}_t\right\|_{\max}$. Then it can be induced that $\dot{B}_{ai}(\boldsymbol{\omega})\geq-k_{a2}B_{ai}(\boldsymbol{\omega})$, so $B_{ai}(\boldsymbol{\omega})\geq0$. Thus, $\boldsymbol{Y}_{ai}\boldsymbol{\omega}\geq-\alpha_a h_{ai}$ holds for all the attitude constraints in $\boldsymbol{H}_a$, which means $h_{ai}\geq0$ always holds. The safety of attitude control can be guaranteed.
\subsubsection{Stability Analysis for Attitude Controller}
In this section, we discuss the stability of the proposed attitude controller. Let $\boldsymbol{\omega}_e=\boldsymbol{\omega}-\boldsymbol{\omega}_c$ be the velocity tracking error. As discussed in Sec.\ref{secpsa}, $\boldsymbol{\omega}_s$ can also be expressed as $\boldsymbol{\omega}_s=\boldsymbol{\omega}_c+\Delta\boldsymbol{\omega}$. 

\emph{Theorem} 3: For the attitude dynamics Eq.(\ref{eq2b}), the attitude control error will converge into a compact set, which is related to the modification effect of the safe angular velocity to the virtual control variable, under the attitude controller Eq.(\ref{eq28}).

\emph{Proof}: The proof is relegated to the Appendix C.

\section{Numerical Simulation}
\label{sec5}
The numerical simulation is conducted to illustrate the effectiveness of the proposed control strategy. The simulations are implemented in MATLAB2022b on PC with Intel Core i5-8400 2.80 GHz and 16GB RAM. The QP problem is solved by using the MATLAB function \emph{quaprog}.
\subsection{Simulation Settings}
In this simulation, the service check eight points in turn for the target, as Table \ref{tab1} shows. The parameters of the service are listed as follows:
\begin{equation*}
	\begin{aligned}
		m&=20{\rm {kg}} \\
		\boldsymbol{J}&=\begin{bmatrix}0.660429&0.014514&0.008125\\0.014514&0.847357&0.035428\\0.008125&0.035428&0.783912\end{bmatrix}{\rm{kg\cdot m^2}}
	\end{aligned}
\end{equation*}

\begin{table}[hbt!]
	\caption{Desired Check Points and Pointing Vectors in Frame $\mathcal{O}$}
	\centering
	\begin{tabular}{lcccccc}
		\hline
		& Number & Check Points(m) & Pointing Vector \\
		\hline
	    &1 & [7 0 0]&[-1 0 0]\\
	    &2 & [0 0 -7]&[0 0 1]\\
	    &3 & [-7 0 0]&[1 0 0]\\
	    &4 & [0 0 7]&[0 0 -1]\\
	    &5 & [0 -11 5]&[0 0 -1]\\
	    &6 & [0.2 -10 -5]&[0 0 1]\\
	    &7 & [0 11 -5]&[0 0 1]\\
	    &8 & [0.2 10 5]&[0 0 -1]\\
		\hline
	\end{tabular}
	\label{tab1}
\end{table}

The controller parameters are set as: $\alpha_a=0.6$, $\alpha_p=0.55$, $\gamma_a=0.001$, $\gamma_p=0.01$, $k_{a1}=0.2$, $k_{a2}=1.1$, $k_{p1}=0.55$, $k_{p2}=0.2$, $\epsilon=0.05$, $\boldsymbol{L}=\operatorname{diag}(\left[0.1,0.1,0.1,0.2,0.2,0.2\right])$, $\boldsymbol{v}_{\max}=0.2$m/s, $\boldsymbol{\omega}_{\max}=2^{\circ}/$s. Besides, the optical axis of the camera is set as $\boldsymbol{P}_{\mathcal{B}3}=\left[1,0,0\right]^T$ and the optical axis of star trackers are set as $\boldsymbol{P}_{\mathcal{B}1}=\left[-\frac{\sqrt{2}}{2},0,-\frac{\sqrt{2}}{2}\right]^T$, $\boldsymbol{P}_{\mathcal{B}2}=\left[-\frac{\sqrt{2}}{2},0,\frac{\sqrt{2}}{2}\right]^T$. The sun exclusive angle is set as $\Psi_{s}=25^{\circ}$ and $\Psi_{c}=30^{\circ}$, and the Earth exclusive angle for star tracker is set as $\Psi_{e}=30^{\circ}$.

The initial orbit elements of the target are listed in Table \ref{tab2}.
\begin{table}[hbt!]
	\caption{Initial Orbit Elements of the Target}
	\centering
	\begin{tabular}{lcccccc}
		\hline
		& Parameter & Value & Unit \\
		\hline
		&Semimajor axis &42139 &km\\
		&Eccentricity &0.002&-\\
		&Inclination &5.3707&deg\\
		&RAAN & 51.2091&deg\\
		&Argument of perigee &236.3791&deg\\
		&Mean anomaly & 59.4097&deg\\
		\hline
	\end{tabular}
	\label{tab2}
\end{table}
Further, the external disturbance force and torque in simulations are set as
\begin{equation*}
	\begin{aligned}
		\boldsymbol{F}_d&=\left[0.01\sin(0.02t),0.02\cos(0.01t),0.01\sin(0.03t)\right]^T\rm{N}\\
		\boldsymbol{T}_d&=\left[0.001\sin(0.03t),0.002\sin(0.02t),0.001\cos(0.03t)\right]^T\rm{Nm}
	\end{aligned}
\end{equation*}
such a disturbance is much bigger than the actual one in space environment, thus it is sufficient for robust evaluation.

The mass and inertia of the service given to the controller exist 20\% uncertainty. The initial relative position of the service is set as $\boldsymbol{r}(0)=\left[15,0,0\right]^T$m, the initial relative velocity is set as $\boldsymbol{v}(0)=\left[0.02,0.01,-0.01\right]^T$m/s. 
\subsection{Simulation Results}
The simulation results are listed as follows. 
\begin{figure}[htb!]
	\centering
	\includegraphics[scale=0.35]{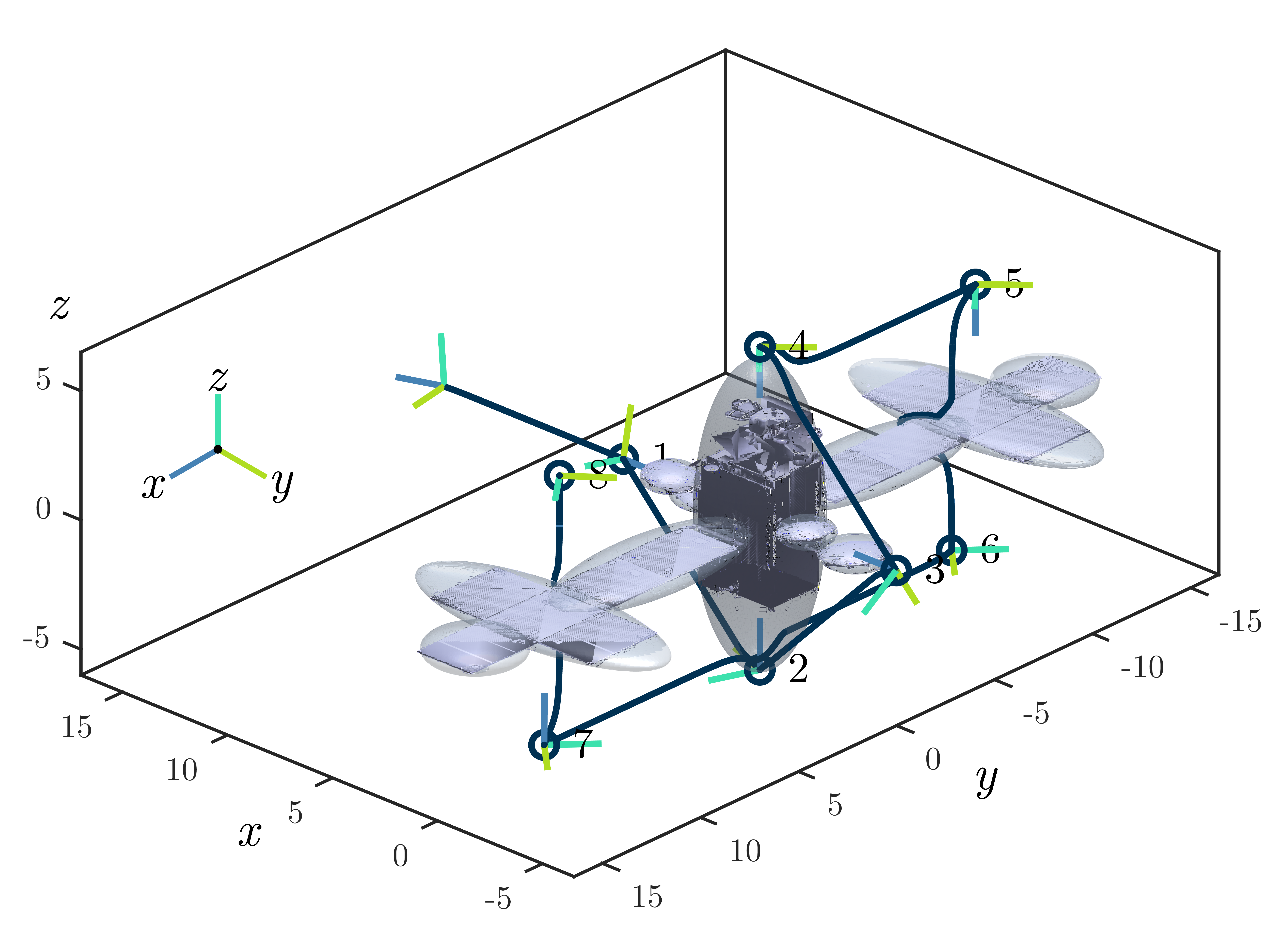}
	\caption{Trajectory of the service in frame $\mathcal{O}$: the blue circles represent the check points}
	\label{fig3d}
\end{figure}
\begin{figure}[htb!]
	\centering
	\includegraphics[scale=0.55]{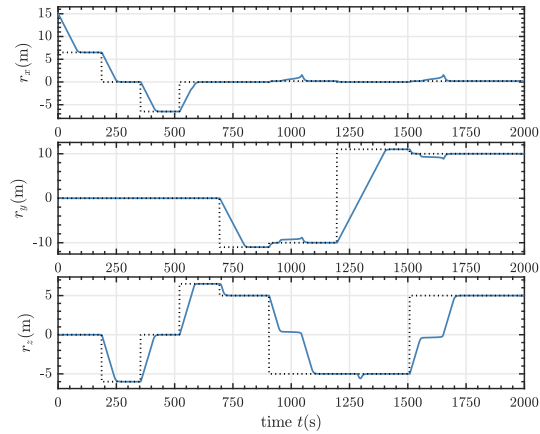}
	\caption{Trajectory of relative position $\boldsymbol{r}$: the black dot dash line represents the target position and the blue line represents the trajectory}
	\label{figr}
\end{figure}

Fig.(\ref{fig3d}) and Fig(\ref{figr}) show the trajectory of the service in frame $\mathcal{O}$, the attitude of the service relative to the target is displayed at the check point. From Fig.(\ref{fig3d}), it can be observed that the service check the desired points and keep attitude to the right direction.
\begin{figure}[htb!]
	\centering
	\includegraphics[scale=0.6]{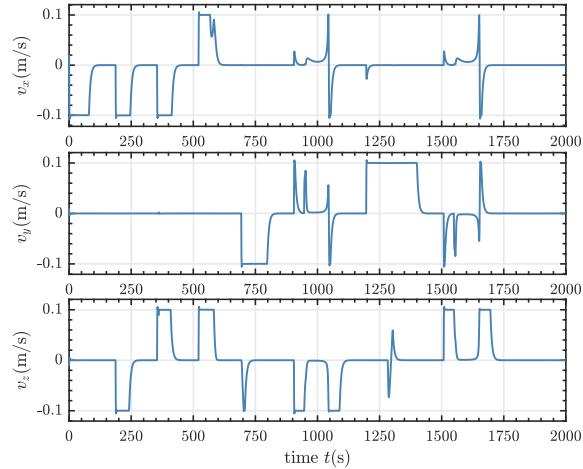}
	\caption{Relative velocity $\boldsymbol{v}$ curves}
	\label{figv}
\end{figure}
\begin{figure}[htb!]
	\centering
	\includegraphics[scale=0.6]{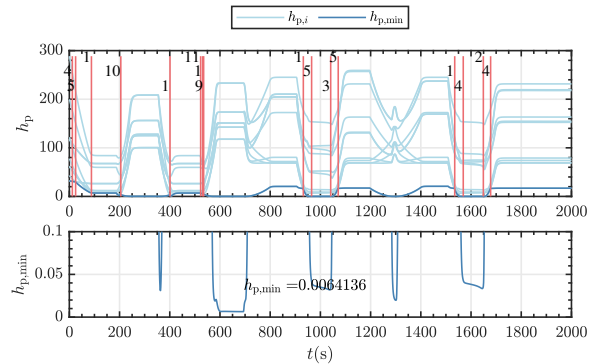}
	\caption{Curves of the value of position CBFs}
	\label{fighv}
\end{figure}

Fig(\ref{figv}) shows the relative velocity of the service and it can be observed that the maximum velocity is constrained in the range of about $\left[-0.1,0.1\right]$m/s. Fig(\ref{fighv}) shows the value of position CBFs: 1) in the subgraph above, the light blue lines represent the value of eleven position CBFs, the solid blue line represents the minimum value of the position CBFs and the red lines represent the number of the current minimum CBF; 2) in the subgraph below, the details of the minimum $h_{pi}$ are shown and the value is 0.0064136, which means that the position constraints are satisfied during the whole inspection mission. 

\begin{figure}[htb!]
	\centering
	\includegraphics[scale=0.6]{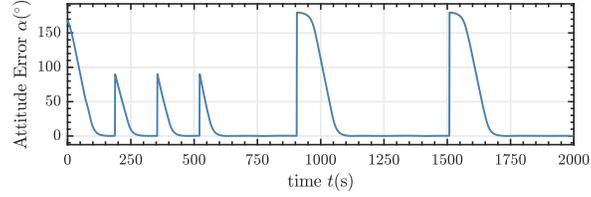}
	\caption{Trajectory of attitude pointing error}
	\label{figerror}
\end{figure}
\begin{figure}[htb!]
	\centering
	\includegraphics[scale=0.6]{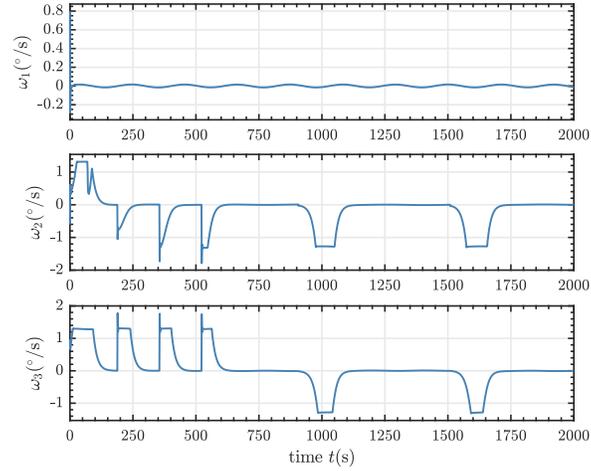}
	\caption{Trajectory of attitude angle velocity}
	\label{figw}
\end{figure}

\begin{figure}[htb!]
	\centering
	\includegraphics[scale=0.6]{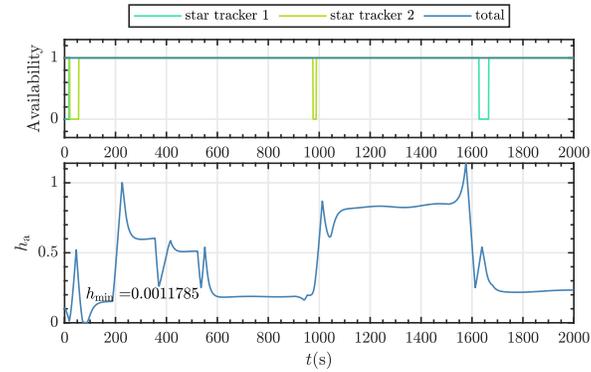}
	\caption{Curses of attitude CBFs}
	\label{fighw}
\end{figure}

Fig(\ref{figerror}) displays the attitude pointing error during the inspection mission. Fig(\ref{figw}) shows the three axis angular velocity of the service and it can be observed that the angular velocity is constrained to be in the range of about $\left[-1.4,1.4\right]^{\circ}/$s.

Fig(\ref{fighw}) gives information about the attitude CBFs: 1) in the subgraph above, the availability of the star trackers is listed and it can be concluded that at least one star tracker is available all the time; 2) in the subgraph below, the details of the minimum $h_{ai}$ are shown and the value is 0.0011785, which means that the attitude constraints are satisfied during the whole inspection mission.

From Fig(\ref{fighv}) and Fig(\ref{fighw}), we can conclude that the position and attitude constraints are satisfied by the proposed controller, which illustrate the effectiveness of the control strategy. Besides, Fig(\ref{figv}) and Fig(\ref{figw}) display that the position and attitude velocity can also be constraint in a certain range, which is smaller than $\boldsymbol{v}_{\max}$ and $\boldsymbol{\omega}_{\max}$. This is because of the term $\alpha_a$ and $\alpha_p$ in the controllers. In fact, the desired safe velocities for attitude and position controller can be regarded as $\frac{k_{a2}}{k_{a2}+\alpha_a}\boldsymbol{\omega}_s$ and $\frac{k_{p2}}{k_{p2}+\alpha_p}\boldsymbol{v}_s$, which means that the designed controller can also restrict the velocities by tracking scaled safe velocities. 
\section{Conclusions}
\label{sec6}
In this paper, we investigate the safety control problem in the spacecraft inspection missions. We propose a control strategy based on control barrier function that makes safety check on kinematics and conducts velocity tracking on dynamics. The safe velocities that satisfy constraints are generated by solving a QP problem and proportional-like position and attitude controllers are designed to tracking safe velocities. Numerical simulations show that the designed control strategy can achieve safe inspection mission and the constraints can be satisfied. Additionally, the numerical simulations show that the proportional-like controllers can guarantee safety despite the velocity tracking error and relax the performance requirements for velocity tracking controllers.

\section*{Appendix A: Proof of Theorem 1}
\label{appa}
\setcounter{equation}{0}
\renewcommand\theequation{A\arabic{equation}} 
Define estimation error $\boldsymbol{e}=\boldsymbol{d}-\hat{\boldsymbol{d}}$. Consider Eq.(\ref{eq16}) and the dynamic of $\boldsymbol{e}$ can be written as
\begin{equation}
	\label{eqA1}
	\begin{aligned}
		\dot{\boldsymbol{e}}&=\dot{\boldsymbol{d}}-\dot{\hat{\boldsymbol{d}}} 
		=\dot{\boldsymbol{d}}-\dot{\boldsymbol{z}}-\frac{\partial \boldsymbol{p}}{\boldsymbol{x}}\dot{\boldsymbol{x}} \\
		&=\dot{\boldsymbol{d}}+\boldsymbol{L}\boldsymbol{z}+\boldsymbol{L}(\boldsymbol{p}-\boldsymbol{d}) 
		=-\boldsymbol{L}\boldsymbol{e}+\dot{\boldsymbol{d}}
	\end{aligned}
\end{equation}

Define a candidate Lyapunov function $V=\frac{1}{2}\boldsymbol{e}^T\boldsymbol{e}$, and take the time derivative of $V$
\begin{equation}
	\label{eqA2}
	\begin{aligned}
		\dot{V}&=\boldsymbol{e}^T\dot{\boldsymbol{e}} 
		=\boldsymbol{e}^T(-\boldsymbol{L}\boldsymbol{e}+\dot{\boldsymbol{d}}) \\
		&\leq -\lambda_{\min}(\boldsymbol{L})\boldsymbol{e}^T\boldsymbol{e}+\left\|\boldsymbol{e} \right\|\delta 
	\end{aligned}
\end{equation} 

Consider the inequality $\left\| \boldsymbol{e}\right\|\delta\leq \frac{\mu}{2}\left\|\boldsymbol{e} \right\|^2+\frac{1}{2\mu}\delta^2 $ and substitute it into Eq.\ref{eqA2}
\begin{equation}
	\label{eqA3}
	\dot{V}\leq -(\lambda_{\min}(\boldsymbol{L})-\frac{\mu}{2})\left\|\boldsymbol{e} \right\|^2+ \frac{1}{2\mu}\delta^2
\end{equation}

According to the \emph{Lemma} 1.2 in \citep{ge2004adaptive}, the estimation error $\boldsymbol{e}$ will converge exponentially to a compact set $\boldsymbol{\Omega}_1=\left\lbrace \boldsymbol{e}|\left\| \boldsymbol{e}\right\| \leq \frac{\delta}{\sqrt{\mu(\lambda_{\min}(\boldsymbol{L})-\frac{\mu}{2})}} \right\rbrace $.
\section{Appendix B: Proof of Theorem 2}
\label{appb}
\setcounter{equation}{0}
\renewcommand\theequation{B\arabic{equation}}
Define the position error Lyapunov function as
\begin{equation}
	\label{eq4}
	V_{p1}=\frac{1}{2}\boldsymbol{r}_e^T\boldsymbol{r}_e
\end{equation}

Taking the time derivative of $V_{p1}$ yields
\begin{equation}
	\label{eq5}
	\dot{V}_{p1}=\boldsymbol{r}_e^T\dot{\boldsymbol{r}}_e=\boldsymbol{r}_e^T\boldsymbol{v}
\end{equation}

Define the candidate Lyapunov function
\begin{equation}
	\label{eq23}
	V_{p2}=V_{p1}+\frac{1}{2}\boldsymbol{v}_e^T\boldsymbol{v}_e
\end{equation}

Taking the time derivative of Eq.(\ref{eq23}) and considering the controller Eq.(\ref{eq6}), Eq.(\ref{eq17}), we have
\begin{equation}
	\label{eq24}
	\begin{aligned}
			\dot{V}_{p2}&=\boldsymbol{r}_e^T\boldsymbol{v}+\boldsymbol{v}_e^T(\dot{\boldsymbol{v}}-\dot{v}_c) \\
			&=\boldsymbol{r}_e^T\left(\boldsymbol{v}_e+\boldsymbol{v}_c\right)+\boldsymbol{v}_e^T\left(-\boldsymbol{C}_o\boldsymbol{r}-\boldsymbol{D}_o\boldsymbol{r}+\boldsymbol{g}+\frac{1}{m}\boldsymbol{F}+\boldsymbol{d}_f+k_{p1}\boldsymbol{v}\right)  \\
			&=-k_{p1}\boldsymbol{r}_e^T\boldsymbol{r}_e+\boldsymbol{r}_e^T\boldsymbol{v}_e+\boldsymbol{v}_e^T\left[-\left(\alpha_p+k_{p2}\right)\boldsymbol{v}+k_{p2}\boldsymbol{v}_c+k_{p2}\Delta\boldsymbol{v}+k_{p1}\boldsymbol{v}+\boldsymbol{d}_f-{\hat{\boldsymbol{d}}}_f\right]
		\end{aligned}
\end{equation}

Let $k_{p1}=\alpha_p$, then we have
\begin{equation}
	\label{eq25}
	\begin{aligned}
			\dot{V}_{p2}&=-k_{p1}\boldsymbol{r}_e^T\boldsymbol{r}_e+\boldsymbol{r}_e^T\boldsymbol{v}_e+\boldsymbol{v}_e^T\left[-k_{p2}\left(\boldsymbol{v}-\boldsymbol{v}_c\right)+k_{p2}\Delta\boldsymbol{v}+\boldsymbol{e}_f\right] \\
			&=-k_{p1}\boldsymbol{r}_e^T\boldsymbol{r}_e+\boldsymbol{r}_e^T\boldsymbol{v}_e-k_{p2}\boldsymbol{v}_e^T\boldsymbol{v}_e+k_{p2}\boldsymbol{v}_e^T\Delta \boldsymbol{v}+\boldsymbol{v}_e^T\boldsymbol{e}_f \\
			&\leq-k_{p1}\left\|\boldsymbol{r}_e\right\|^2-k_{p2}\left\|\boldsymbol{v}_e\right\|^2+k_{p2}\left\|\boldsymbol{v}_e\right\|\left\|\Delta\boldsymbol{v}\right\|+\left\|\boldsymbol{v}_e\right\|\left\|\boldsymbol{e}_f\right\|+\left\|\boldsymbol{v}_e\right\|\left\|\boldsymbol{r}_e\right\|   
		\end{aligned}
\end{equation}

Let $\Theta_p=k_{p2}\left\|\Delta\boldsymbol{v}\right\|+\left\|\boldsymbol{e}_f\right\|$, and consider the following inequalities
\begin{equation}
	\label{eq26}
	\begin{aligned}
			\left\|\boldsymbol{v}_e\right\|\Theta_p&\leq \frac{\kappa_1}{2}\left\|\boldsymbol{v}_e\right\|^2+\frac{1}{2\kappa_1}\Theta_p^2 \\
			\left\|\boldsymbol{v}_e\right\| \left\|\boldsymbol{r}_e\right\|&\leq \frac{1}{2}\left\|\boldsymbol{v}_e\right\|^2+\frac{1}{2}\left\|\boldsymbol{r}_e\right\|^2
		\end{aligned}
\end{equation}

Substituting Eq.(\ref{eq26}) into Eq.(\ref{eq25}) yields
\begin{equation}
	\label{eq27}
	\begin{aligned}
			\dot{V}_{p2}&\leq-\left(k_{p1}-\frac{1}{2}\right)\left\|\boldsymbol{r}_e\right\|^2-\left(k_{p2}-\frac{\kappa_1}{2}-\frac{1}{2}\right)\left\|\boldsymbol{v}_e\right\|^2+\frac{1}{2\kappa_1}\Theta_p^2 \\
			&= -\chi_p V_{p2}+\Lambda_p 
		\end{aligned}
\end{equation}
where $\kappa_1$ is a positive constant to be designed, $\chi_p=\min\left\lbrace 2k_{p1}-1, 2k_{p2}-\kappa-1\right\rbrace$, $\Lambda_p=\frac{1}{2\kappa}\Theta_p^2$. Let $k_{p1}=\alpha_p>\frac{1}{2}$, $k_{p2}>\frac{\kappa_1+1}{2}$, then the position control error will converge exponentially to a compact set $\Omega_2 =\left\lbrace\boldsymbol{x}_{pe}|\left\|\boldsymbol{x}_{pe}\right\|\leq\sqrt{\frac{2\Lambda_p}{\chi_p}}\right\rbrace$, where $\boldsymbol{x}_{pe}=\left[\boldsymbol{r}_e^T,\boldsymbol{v}_e^T\right]^T$.
\section{Appendix C: Proof of Theorem 3}
\label{appd}
\setcounter{equation}{0}
\renewcommand\theequation{C\arabic{equation}}
Let $\boldsymbol{s}=\boldsymbol{P}_{\mathcal{B}3}-\boldsymbol{\Gamma}$ be the attitude error vector and define the following candidate attitude Lyapunov function

\begin{equation}
	\label{eq12}
	\begin{aligned}
		V_{a1}=\frac{1}{2}\boldsymbol{s}^T\boldsymbol{s} 
		=1-\boldsymbol{P}_{\mathcal{B}3}^T\boldsymbol{\Gamma}
	\end{aligned}
\end{equation}

Taking the time derivative of $V_{a1}$ yields
\begin{equation}
	\label{eq13}
	\dot{V}_{a1}=-\boldsymbol{P}_{\mathcal{B}3}^T\left(\boldsymbol{\Gamma}\times\boldsymbol{\omega} \right) =-\boldsymbol{P}_{\mathcal{B}3}^T\boldsymbol{\Gamma}^{\times}\boldsymbol{\omega}
\end{equation}

Define the candidate Lyapunov function as
\begin{equation}
	\label{eq35}
	V_{a2}=V_{a1}+\frac{1}{2}\boldsymbol{\omega}_e^T\boldsymbol{\omega}_e
\end{equation}

It can be induced that $\left\|\boldsymbol{\omega}_c\right\|=k_{a1}\left\|\boldsymbol{P}_{\mathcal{B}3}\right\|\left\|\boldsymbol{\Gamma}\right\|\sin \alpha=k_{a1}\sin \alpha\leq k_{a1}$, where $\alpha$ is the angle between $\boldsymbol{P}_{\mathcal{B}3}$ and $\boldsymbol{\Gamma}$. Similarly, $\left\|\boldsymbol{\Phi}\right\|=\sin \alpha\leq1$. Taking the time derivative of Eq.(\ref{eq35}) and considering the controller Eq.(\ref{eq14}), Eq.(\ref{eq28}), we have
\begin{equation}
	\label{eq36}
	\begin{aligned}
		\dot{V}_{a2}&=-\boldsymbol{\Phi}^T\boldsymbol{\omega}+\boldsymbol{\omega}_e^T\left(\dot{\boldsymbol{\omega}}-\dot{\boldsymbol{\omega}}_c\right)\\
		&=-\boldsymbol{\Phi}^T(\boldsymbol{\omega}_e+\boldsymbol{\omega}_c)+\boldsymbol{\omega}_e^T\left\lbrace \boldsymbol{J}^{-1}\left[ -{\boldsymbol{\omega}}\times(\boldsymbol{J}{\boldsymbol{\omega}})+\boldsymbol{T}\right] +\boldsymbol{d}_t-k_{a1}\boldsymbol{P}_{\mathcal{B}3}^{\times}\boldsymbol{\Gamma}^{\times}\boldsymbol{\omega}\right\rbrace \\
		&=-k_{a1}\boldsymbol{\Phi}^T\boldsymbol{\Phi}-\boldsymbol{\Phi}^T\boldsymbol{\omega}_e+\boldsymbol{\omega}_e^T\left[-(\alpha_a+k_{a2})\boldsymbol{\omega}+k_{a2}(\boldsymbol{\omega}_c+\Delta\boldsymbol{\omega})+\boldsymbol{e}_t-k_{a1}\boldsymbol{P}_{\mathcal{B}3}^{\times}\boldsymbol{\Gamma}^{\times}\boldsymbol{\omega}\right] \\
		&=-k_{a1}\boldsymbol{\Phi}^T\boldsymbol{\Phi}-k_{a2}\boldsymbol{\omega}_e^T\boldsymbol{\omega}_e+\boldsymbol{\omega}_e^T(-\boldsymbol{\Phi}-\alpha_a\boldsymbol{\omega}+k_{a2}\Delta\boldsymbol{\omega}+\boldsymbol{e}_t-k_{a1}\boldsymbol{P}_{\mathcal{B}3}^{\times}\boldsymbol{\Gamma}^{\times}\boldsymbol{\omega}) \\
		&\leq-k_{a1}\boldsymbol{\Phi}^T\boldsymbol{\Phi}-k_{a2}\boldsymbol{\omega}_e^T\boldsymbol{\omega}_e+\left\|\boldsymbol{\omega}_e^T\right\|\left\|\boldsymbol{\Phi}\right\|+k_{a2}\left\|\boldsymbol{\omega}_e^T\right\|\left\|\Delta\boldsymbol{\omega}\right\|+\left\|\boldsymbol{\omega}_e^T\right\|\left\|\boldsymbol{e}_t\right\|\\
		&\quad +(\alpha_a+k_{a1})\left\|\boldsymbol{\omega}_e^T\right\|\left\|\boldsymbol{\omega}_c\right\|-\alpha_a\boldsymbol{\omega}_e^T\boldsymbol{\omega}_e
	\end{aligned}
\end{equation}

Considering following inequalities
\begin{equation}
	\label{eq37}
	\begin{aligned}
		\left\|\boldsymbol{\omega}_e^T\right\|\left\|\boldsymbol{\omega}_c\right\|&\leq k_{a1}\left\|\boldsymbol{\omega}_e\right\|\leq\frac{k_{a1}^2}{2\kappa_2}+\frac{\kappa_2}{2}\left\|\boldsymbol{\omega}_e\right\|^2 \\
		\left\|\boldsymbol{\omega}_e^T\right\|\left\|\boldsymbol{\Phi}\right\|&\leq\frac{1}{2}\left\|\boldsymbol{\omega}_e\right\|^2+\frac{1}{2}\left\|\boldsymbol{\Phi}\right\|^2\leq\frac{1}{2}\left\|\boldsymbol{\omega}_e\right\|^2+\frac{1}{2}
	\end{aligned}
\end{equation}

Let ${\Theta}_a=k_{a2}\left\|\Delta\boldsymbol{\omega}\right\|+\left\|\boldsymbol{e}_t\right\|$, then we have
\begin{equation}
	\label{eq38}
	\begin{aligned}
		\left\|\boldsymbol{\omega}_e\right\|{\Theta}_a&\leq\frac{1}{2}\left\|\boldsymbol{\omega}_e\right\|^2+\frac{1}{2}{\Theta}_a^2
	\end{aligned}
\end{equation}

Substituting Eq.(\ref{eq37}) and Eq.(\ref{eq38}) into Eq.(\ref{eq36}) yields
\begin{equation}
	\label{eq39}
	\begin{aligned}
		\dot{V}_{a2}\leq-k_{a1}\left\|\boldsymbol{\Phi}\right\|^2-\left(k_{a2}+\alpha_a-k_{a1}-1-\frac{\kappa_2(\alpha_a+k_{a1})}{2}\right)\left\|\boldsymbol{\omega}_e\right\|^2+\frac{k_{a1}^2(\alpha_a+k_{a1})}{2\kappa_2}+\frac{1}{2}+\frac{1}{2}{\Theta}_a^2
	\end{aligned}
\end{equation}

Considering that
\begin{equation}
	\label{eq40}
	\begin{aligned}
		\left\|\boldsymbol{\Phi}\right\|^2=\sin^2\alpha=1-\cos^2\alpha=(1+\cos\alpha)(1-\cos\alpha)=(1+\cos\alpha)(1-\boldsymbol{P}_{\mathcal{B}3}^T\boldsymbol{\Gamma})=(1+\cos\alpha)V_{a1}
	\end{aligned}
\end{equation}

Then we have
\begin{equation}
	\label{eq41}
	\begin{aligned}
		\dot{V}_{a2}\leq-k_{a1}(1+\cos\alpha)V_{a1}-\left(k_{a2}+\alpha_a-k_{a1}-1-\frac{\kappa_2(\alpha_a+k_{a1})}{2}\right)\left\|\boldsymbol{\omega}_e\right\|^2+\frac{k_{a1}^2(\alpha_a+k_{a1})}{2\kappa_2}+\frac{1}{2}{\Theta}_a^2+\frac{1}{2} 
	\end{aligned}
\end{equation}

Let $k_{a1}>0$ and select proper $\kappa_2$ and $\alpha_a$ to make $k_{a2}>-\alpha_a+k_{a1}+1+\frac{\kappa_2(\alpha_a+k_{a1})}{2}$, then Eq.(\ref{eq41}) will be
\begin{equation}
	\label{eq42}
	\dot{V}_{a2}\leq -\chi_a V_{a2}+\Lambda_a
\end{equation}
where
\begin{equation*}
	\chi_a=\min\left\lbrace k_{a1}(1+\cos\alpha),2k_{a2}+2\alpha_a-2k_{a1}-2-\kappa_2(\alpha_a+k_{a1})\right\rbrace
\end{equation*}
\begin{equation*}
	\Lambda_a=\frac{k_{a1}^2(\alpha_a+k_{a1})}{2\kappa_2}+\frac{1}{2}{\Theta}_a^2+\frac{1}{2}
\end{equation*}

Then the attitude control error will converge exponentially to a compact set $\Omega_3=\left\lbrace\boldsymbol{x}_{ae}|\left\|\boldsymbol{x}\right\|_{ae}\leq\sqrt{\frac{2\Lambda_a}{\chi_a}}\right\rbrace$, where $\boldsymbol{x}_{ae}=\left[\boldsymbol{s}^T,\boldsymbol{\omega}_e^T\right]^T$.

\bibliography{wk-refs}

\end{document}